# Variability within Modeling Language Definitions

María Victoria Cengarle[1], Hans Grönniger[2], and Bernhard Rumpe[2]

[1] Software and Systems Engineering, Technische Universität München, Germany
[2] Lehrstuhl Software Engineering, RWTH Aachen, Germany

**Abstract.** We present a taxonomy of the variability mechanisms offered by modeling languages. The definition of a formal language encompasses a syntax and a semantic domain as well as the mapping that relates them, thus language variabilities are classified according to which of those three pillars they address. This work furthermore proposes a framework to explicitly document and manage the variation points and their corresponding variants of a variable modeling language. The framework enables the systematic study of various kinds of variabilities and their interdependencies. Moreover, it allows a methodical customization of a language, for example, to a given application domain. The taxonomy of variability is explicitly of interest for the UML to provide a more precise understanding of its variation points.

**Keywords:** Modeling languages, variability, formal semantics, UML.

## 1 Introduction

A complete definition of a formal modeling language consists of the description of its syntax and its semantics (meaning) [1]. It is widely accepted that a commonly agreed formal definition (especially semantics) of a language helps to avoid misunderstandings and lack of interoperability between tools.

In [2], we presented a tool-based approach to define textual modeling languages and to formalize their semantics in a flexible way using a theorem prover. While one of our main targets is the formalization of the Unified Modeling Language (UML 2) [3,4], the approach is more general and applies to any modeling language based on objects.

In this paper, we investigate how variability in a language definition can be formally specified. This work is inspired by the introduction of semantic variation points in UML where portions of the language have been deliberately incompletely specified. The benefits of systematically describing UML's variability have been noted early [5]. The treatment of semantic variation points in the UML, however, is rather disappointing. It was not systematically carried out, semantic variation points are dispersed across the documentation. Moreover, the standard fails to tag them completely: it suffices to look for underspecified semantic definitions in order to realize that there are far more semantic variation points than those explicitly labeled as such.

Beyond UML, we are interested in a general treatment of variability in modeling languages which may be of semantic and also of syntactic nature. Hence, one goal of this work is to classify the kinds of variability that a modeling language may offer and their interdependencies. Additionally, we extend our approach from [2] and present a tool-based solution to define and configure variability within a language definition.

A systematic approach to variability should make it possible to explicitly state all (possibly implicit) assumptions and previously chosen variants. This allows a systematic customization of a language for a given application domain. Furthermore, tool builders can refer to particular variants in order to document design decisions. Variation points of modeling languages, unlike those of product lines, are not associated with a binding time [6]. That is, tool builders may delay the binding of a variation point to a variant and leave the decision to project managers. Moreover, these may even forward the disambiguation to modelers. As for UML, currently *implementors may provide [...] informal feature support statements [...] for less precisely defined dimensions such as presentation options and semantic variation points"* [3, Sect. 2.3]. We improve this situation by making precise the definition of the variability mechanisms offered by a language.

The rest of this paper is organized as follows. Sect. 2 describes the constituents of a modeling language definition. Sect. 3 presents our classification of variability in a language definition. Sect. 4 introduces our tool-supported solution using feature diagrams. The approach is illustrated with a simple example of UML-like class diagrams. Sect. 5 discusses related work and Sect. 6 draws conclusions and sketches future work.

## 2  Constituents of a Modeling Language Definition

As shown in Fig. 1, a complete definition of a modeling language consists of the following basic parts:

- the concrete syntax of the language, which may be a graphical or textual syntax or a combination of both,
- the abstract syntax to which the concrete syntax is mapped. For a textual syntax this may be given as abstract syntax trees. In case of graphical modeling, metamodels are typically used. Additionally, a set of well-formedness rules or context conditions are defined,
- some minimal abstract syntax that can be derived from the abstract syntax by expressing more complex constructs of the language by primitive ones. Thereby the number of constructs but not the expressive power of the language is reduced. This eases the definition of the semantics of the language. This step may not be required for some languages,
- a semantic domain, a domain well-known and understood, typically based on a well-defined mathematical theory, and
- the semantic mapping that relates elements of the (minimal) abstract syntax to elements of the semantic domain.

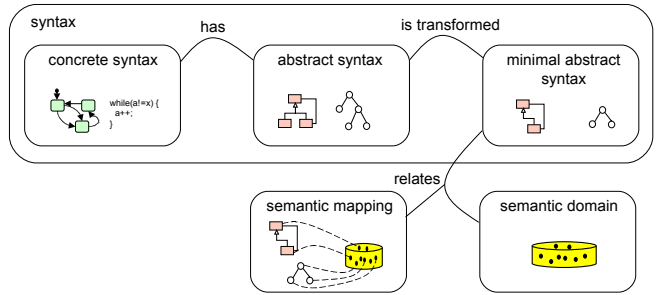

**Fig. 1.** Basic parts of a modeling language definition

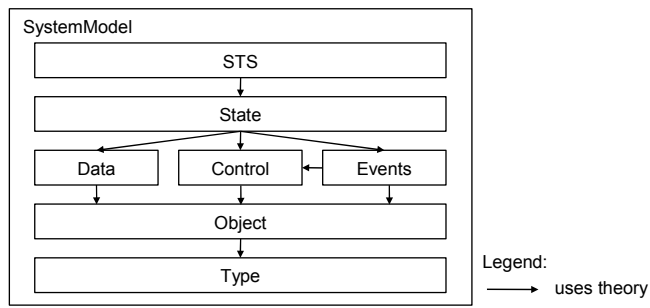

**Fig. 2.** Theories that constitute the system model

Characteristic for our approach to define the semantics of a modeling language is a set-valued or predicative semantic mapping of the form $sem(.) : \mathcal{L} \to \wp(\mathcal{S})$. The semantics of a model as an element of the (minimal) abstract syntax $m \in \mathcal{L}$ is therefore the set $sem(m)$ of elements in the semantic domain $\mathcal{S}$.

We defined a single semantic domain $\mathcal{S}$ used as a target for the semantic mapping of various kinds of object-oriented modeling languages [7]. This domain, called system model, captures and integrates all aspects of object-oriented systems using basic mathematical theories. It is rather detailed as it defines various structural, behavioral, and interaction aspects, and is built in a modular fashion as depicted in Fig. 2. Systems in the system model are state transition systems (theory STS). They operate on a global system state which is composed of object individual states (theory State). States constitute a data store for attribute values of objects (theory Data), a control store (theory Control) for active threads and computational states of methods, and an event store for unprocessed events (theory Events). States evolve dynamically. Static information (e.g., which classes, methods, etc., exist) is defined through underspecified universes containing abstract identifiers only. For example, UTYPE is the universe of type names (defined in theory Type). Classes are elements of the universe UCLASS (theory Object) and are only described by functions that yield information about their attributes or methods, i.e., they are not constructed from records. Thus, the definition of the system model is predicative and not constructive. For a complete picture of the system model features, the reader is referred to [7].

The system model as a single semantic domain and the set-valued semantic mapping enable a straightforward treatment of composition and refinement of possibly incomplete and underspecified models of various modeling languages [8]. For example, the integrated semantics of models $m_1, \ldots, m_n$ from possibly different languages $\mathcal{L}_1, \ldots, \mathcal{L}_n$ is given as $sem_{\mathcal{L}_1}(m_1) \cap \ldots \cap sem_{\mathcal{L}_n}(m_n)$. In the same way, $m' \in \mathcal{L}$ is a refinement of $m \in \mathcal{L}$, exactly if $sem(m') \subseteq sem(m)$.

## 3   Classification of Variability

In this section, we develop a classification of variability that a modeling language may offer. We do not restrict our attention to semantic variability (in UML terms, semantic variation points) but also consider syntactic variability.

In a very abstract view, the syntax of a formal language is defined by a set of words over some alphabet $A$, i.e., $\mathcal{L} \subseteq A^*$. *Syntactic variability* allows for defining more than one syntax, say $\mathcal{L}_1$ and $\mathcal{L}_2$, which normally contain many common words but are different. That is, there is at least one model (i.e., word) $m \in (\mathcal{L}_1 \cup \mathcal{L}_2) \setminus (\mathcal{L}_1 \cap \mathcal{L}_2)$ that is in one but not both languages. The semantics of a syntax $\mathcal{L}$ over some semantic domain $\mathcal{S}$ can be defined as $sem \subseteq \mathcal{L} \times \mathcal{S}$ (in a relational style). *Semantic variability* means more than one semantics, say $sem_1$ and $sem_2$, for a given syntax $\mathcal{L}$. These mappings may have different codomains $\mathcal{S}_1 \neq \mathcal{S}_2$ or not. As with the syntax, $sem_1$ and $sem_2$ are mostly the same but there is at least one model $m$ and an element $s$ for which $(m, s) \in (sem_1 \cup sem_2) \setminus (sem_1 \cap sem_2)$. So the meaning of the model differs according to which semantics is chosen.

There naturally may be languages containing both kinds of variability, and relationships between both exist. In the following, we concretize this abstract view by analyzing how variants and their interdependencies can be classified.

### 3.1   Syntactic Variability

Regarding concrete syntax (see Fig. 1), differences can be given by, e.g., alternative keywords such as "public" or "+" in case of modifiers, or the font size, line thickness, and color of some graphical element. In UML, these are called *presentation options* and can be classified as *presentation variability*. They improve the readability of models. Nevertheless, presentation options are so defined that the abstract syntax of models remains the same even if the options are changed.[1]

---

[1] This is an important assumption we make on presentation options, namely that they do *not* alter abstract syntax and hence the intended semantics of the presented model element. Font size, for instance, may have a meaning in cartography, where cities with bigger labels have more inhabitants. In the case of cartography, therefore, font size does matter and is not a presentation option.

We do not classify presentation options as syntactic variability since they do not make it possible to define different languages. Their effect exclusively concerns the concrete syntax. They must, nevertheless, be registered and documented.

The syntax of a language may allow the use of *stereotypes*. The term stereotype, borrowed from UML, is used here to designate a general principle of extending the syntax of a language. The concrete set of defined stereotypes (e.g., as part of a profile in case of UML) is classified as syntactic variability.

Another kind of syntactic variability also found in the syntax is given by so-called *language parameters*. Concerning for instance UML, the language of state machines defines transition systems whose transitions are triggered by a stimulus subject to a condition on the stimulus and/or the internal state of the object. The language in which conditions (or guards) are expressed is not specified. This constitutes a syntactic variability.

In the abstract syntax, *optional context conditions* may exist. Examples thereof, for instance for a particular code generator to operate, are the enforcement of types of attributes of a class to be defined, and the restriction to single inheritance only. Context conditions rule out certain models based on syntactic criteria. Only if the context conditions are met, the model is *well-formed* and it makes sense to give the model a semantics.

The syntax also may offer constructs that enhance readability and are semantically equivalent to other, usually more involved, expressions of the language. Such constructs are often referred to as "syntactic sugar" and may be safely omitted, since models of the language obtained by the use of those constructs can be replaced by equivalent models that do not use the *abbreviations*. We classify this as presentation variability. In particular, the language can be reduced to a minimal one, which not necessarily is unique. Note that a minimal language derived this way may still allow synonyms, i.e., syntactically different models $m_1$ and $m_2$ that denote the same semantics $sem(m_1) = sem(m_2)$.

Summarizing, we classify any variability as syntactic variability that still may be present in the minimal abstract syntax of a modeling language and hence interacts with the semantics. This variability originates from stereotypes, language parameters, and optional context conditions.

### 3.2   Semantic Variability

While UML only uses the term semantic variation point, we further subdivide semantic variability into *semantic mapping variability* and *semantic domain variability*; cf. Fig. 1. A helpful analogy might be to see the variability of the semantic mapping similar to configuration options of a code generator while variability of the semantic domain has its analogy with properties of an underlying run-time system or target platform.

Regarding semantic domain variability, the system model defined in [7] already contains explicit variability in form of extensions through optional definitions. In general, semantic domain variants may provide alternative realizations of functions, additional constraints to properties of existing definitions, or optional structures and definitions. Alternative realizations are, for example, different notions of type-safe method overriding. Additional constraints are, for example, the restriction to single inheritance only, or the requirement of certain predefined types like, e.g., "String."

Similarly, in the semantic mapping, the same mechanisms to introduce variants apply. Semantic mapping variability often manifests as alternative choices for specific mapping functions while the target domain remains the same. For instance, one mapping of super-classes of classes in a UML class diagram assumes multiple inheritance in the semantic domain, while an alternative mapping uses some delegation mechanism for a domain that may lack multiple inheritance. As this example shows, there are also various relationships between variants on the different levels which will be discussed in more detail in the following. As another example, consider the representation of states of a state machine in an implementation as, e.g., a simple enumeration or using the state pattern [9].

Note that semantic variability is transparent to the modeler. But it may be necessary to allow the modeler to select one or the other interpretation of a construct. We propose to model these interpretation choices as syntactic variability by providing corresponding stereotypes. For instance, consider the example of a semantic mapping for a class which states that only a single instance of that class may exist at run-time. One possibility would be to encode this syntactically as a stereotype "singleton" which can be used by the modeler and which is used by the semantic mapping to associate exactly this meaning to the given class.

Table 1 provides a comprehensive summary of our modeling language variability classification.

**Table 1.** Variability classification summary

| | |
|---|---|
| **presentation variability** | variability not present in a minimal abstract syntax |
| presentation options | affect concrete syntax only |
| abbreviations | can be omitted without losing expressiveness |
| **syntactic variability** | variability affecting a minimal abstract syntax |
| stereotypes | syntactic encoding of semantic variability |
| language parameters | usable with different independent languages |
| context conditions | constrain the set of well-formed models |
| **semantic variability** | variability in the semantics |
| semantic domain variability | variability in the underlying target domain |
| semantic mapping variability | different choices for mapping functions |

### 3.3  Interdependency and Consistency

Dependencies between variants exist. These are characterized with the help of examples. Consider the integration of multiple languages: One language might be parameter to another, e.g., a constraint or action language. Additionally, languages may be mainly orthogonal and used to describe different views of the same system such a class and state machine diagrams. In any of these cases, different assumption on the underlying domain may be made, i.e., different variants of the semantic domain may be assumed. Moreover, a language that is parameter to another is equipped with a semantics that has to fit the assumptions made by the parametric language.

Context conditions may influence the selection of a specific semantic mapping. For instance, if the context conditions for UML class diagrams guarantee that multiple inheritance is syntactically excluded, then one can safely select a semantic mapping that only handles single inheritance. Similarly, if a semantic domain only allows for single inheritance, then a delegate mechanism must be resorted to by the semantic mapping of UML class diagrams in case multiple inheritance is allowed syntactically.

From these examples we conclude that it is important to capture all possible variants and their interdependencies. We propose to model them using feature diagrams including constraints that state inclusion or exclusion between variants [10].[2] As a supplement, informal descriptions of the variabilities can be given to explain their raison d'être. The proposed approach is completely supported by tools and will be described in the next section.

Unfortunately, capturing variants as feature diagrams and constraints does not guarantee that a concrete configuration of variants that conforms to the given feature diagrams is consistent. Since we have many configuration options, we might have not captured all constraints to rule out inconsistent, unwanted, or simply uninteresting configurations. Especially when integrating multiple languages, there is a possible risk of contradicting mapping functions. One way to obtain a consistent set of theories is to actually prove consistency. That is, given two languages $\mathcal{L}_1$ and $\mathcal{L}_2$ with semantic mappings $sem_1$ and $sem_2$, to show

$$sem_1(m_1) \cap sem_2(m_2) \neq \emptyset$$

for some witnesses $m_1 \in \mathcal{L}_1$ and $m_2 \in \mathcal{L}_2$.

## 4   Definition and Configuration of Variability

We now describe the actual definition and configuration of variability in a modeling language with respect to the configurable semantic mapping and the likewise configurable semantic domain. Syntactic variability such as optional context conditions and language parameters can be handled similarly and are therefore omitted here. The presentation is accompanied by a simple running example.

The whole approach of defining a language and its variabilities is supported by two tools. The basic tool-based approach (neglecting variability) has been presented in [2] and is summarized below. It features a complete, formal, flexible, and machine-readable definition of modeling languages using the tools MontiCore and Isabelle/HOL.

---

[2] There is an inclusion relation between two or more variants if the choice of one makes it mandatory to choose the other(s). There is an exclusion relation between two or more variants if the choice of one forbids the choice of the other(s).

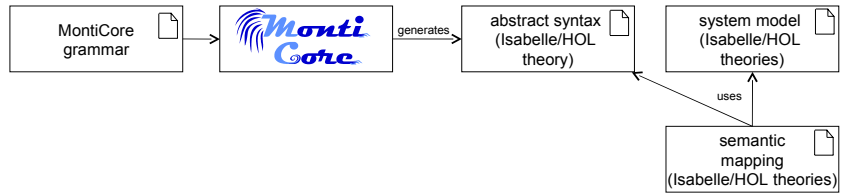

**Fig. 3.** Approach with tool support

### 4.1 Prerequisites

The basic approach is depicted in Fig. 3. MontiCore [11] is a framework for the textual definition of languages based on an extended context-free grammar format. We use MontiCore to define the concrete syntax of a language because it provides enhanced modularity concepts like language inheritance and embedding (not used in the simple running example, though). Sophisticated framework functionality allows, for example, an easy development of generators. Note that the general idea can similarly be implemented using, e.g., metamodeling.

To provide a semantics developer with maximum flexibility and also with some machine checking (e.g., type checking) as well as the potential for verification applications, we use the theorem prover Isabelle/HOL [12] for

- the formalization of the system model as a hierarchy of theories,
- the representation of the abstract syntax of the language as a deep embedding [13], and
- the actual semantic mapping that uses the generated abstract syntax and maps each language construct to predicates over systems of the formalized system model.

**Concrete Syntax.** The example grammar `CDSimp` in Fig. 4 defines UML-like class diagrams with classes that can have super-classes. MontiCore grammars

```
1   grammar CDSimp {
2     CDDefinition = "classdiagram" Name:IDENT "{" (CDClass)* "}";
3
4     CDClass =
5       "class" Name:IDENT ("extends" scl:IDENT ("," scl:IDENT)*)?";";
6   }
```

**Fig. 4.** MontiCore grammar of class diagrams

have terminal symbols enclosed in quotes (see, e.g., Fig. 4, line 2) and support Kleene closure (∗) and option (?), among other constructs. The two rules of `CDSimp` use the built-in identifier rule `IDENT`. Nonterminals may be prefixed by descriptive names followed by a colon (like `IDENT`, l. 2). According to Fig. 4, a class diagram definition (l. 2) has a name and a set of classes. Classes (l. 4) have a name and a comma separated list of names that refer to super-classes.

**Abstract Syntax.** A MontiCore generator produces the Isabelle/HOL data type definition in theory `CDSimpAS` (see Fig. 5) from the grammar in Fig. 4.

```
1   theory CDSimpAS imports GeneralAS
2   begin
3   datatype CDClass =
4       CDClass IDENT "IDENT list"
5
6   datatype CDDefinition =
7       CDDefinition IDENT "CDClass list"
8   end
```

**Fig. 5.** Generated abstract syntax data type in Isabelle/HOL

Isabelle/HOL data types have a name (e.g., `CDClass` in Fig. 5, l. 3), a constructor (also `CDClass`, l. 4), and a list of arguments. Data type `IDENT` is defined in the imported, re-usable theory `GeneralAS` and iteration in a grammar is translated to the built-in data type `list` (e.g., l. 4). A complete account on the mapping of MontiCore grammars to Isabelle/HOL can be found in [2].

**System Model.** We have formalized the system model, introduced in Sect. 2, in Isabelle/HOL as a hierarchy of theories.

```
1   theory Object imports Type
2   begin
3   datatype iCLASS = Class "char list"
4
5   consts
6    UCLASS :: "SystemModel ⇒ iCLASS set"
7    sub :: "SystemModel ⇒ iCLASS ⇒ iCLASS ⇒ bool"
8
9   fun psubRefl :: "SystemModel ⇒ bool"
10    where "psubRefl sm = (∀ C ∈ UCLASS sm . sub sm C C)"
11  end
```

**Fig. 6.** Isabelle/HOL theory `Object` (excerpt)

Fig. 6 shows a small excerpt from the theory `Object` which introduces the universe of classes `UCLASS` (line 6) as a function that yields a set of class names (of type `iCLASS`). `consts` is Isabelle's way of declaring a constant without defining it. Additionally, a subclassing relation `sub` is declared. The boolean function definition `psubRefl` is a simple example of a predicate that must hold in all valid systems and requires reflexivity of the subclassing relation.

The top-level theory `SystemModel-base` (Fig. 7) imports all basic definitions and defines a predicate `valid-base`. In our abbreviated example, only theory `Object` is imported. The full theory would import all other theories from Fig. 2 and combine all predicates (like `psubRefl`) into `valid-base`, describing all properties of a valid system in the system model.

```
1  theory SystemModel-base imports Object
2  begin
3  fun valid-base :: "SystemModel ⇒ bool"
4    where "valid-base sm = (psubRefl sm ∧ ... )"
5  end
```

**Fig. 7.** Isabelle/HOL theory `SystemModel-base` (excerpt)

**Semantic Mapping.** The semantic mapping of our simplified class diagrams is likewise formalized in Isabelle/HOL. The theory in Fig. 8 imports the abstract syntax and the system model theory and defines the mapping. We only state the signatures of the mapping functions, which are built in a modular fashion along the abstract syntax. Note that the mapping functions for classes and class diagrams, `mCDClass` and `mCDDefinition`, are function definitions (using the keyword `fun`) while the mapping of super-classes of a class, `consts mSuperClasses`, again is just a function declaration whose body has not yet been defined.

```
1  theory CDSimpSem-base imports CDSimpAS SystemModel
2  begin
3  consts mSuperClasses :: "iCLASS ⇒ IDENT list ⇒ SystemModel ⇒ bool"
4
5  fun mCDClass :: "CDClass ⇒ SystemModel ⇒ bool"
6    where ...
7
8  fun mCDDefinition :: "CDDefinition ⇒ SystemModel set"
9    where ...
10 end
```

**Fig. 8.** Semantic mapping of the simplified class diagram in Isabelle/HOL

### 4.2 Definition of Variants

We start by introducing a variant for the system model. Fig. 9 contains a theory with an additional constraint for the transitive subclassing relation, restricting it to single inheritance. That is, for all classes `C1`, `C2`, `C3`, if `C1` is a sub class of `C2` and `C3`, then `C2` and `C3` have to be in a subclass relationship (or equal due to reflexivity of `sub`).

As explained before, we model variants of theories as feature diagrams like the one in Fig. 10[3]. Ignoring the check mark for a moment, the feature diagram therein states that `SingleInheritance` is an optional feature of the theory `Object`. Other variants may be associated with other theories as the other variation point `vType` indicates.

---

[3] In our tool suite, we use a textual version of feature diagrams and configuration files but we stick to the standard graphical form for the sake of clarity of the presentation.

```
1  theory SingleInheritance imports Object
2  begin
3  fun valid-SingleInheritance :: "SystemModel ⇒ bool"
4    where "valid-SingleInheritance sm = (∀ C1 C2 C3.
5        sub sm C1 C2 ∧ sub sm C1 C3 ⟶ (sub sm C2 C3 ∨ sub sm C3 C2))"
6  end
```

**Fig. 9.** Definition of an Isabelle/HOL predicate about single inheritance

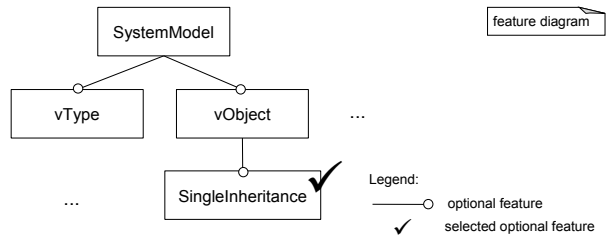

**Fig. 10.** Semantic domain feature diagram (fragment)

Additionally, the feature diagram for the variants of the semantic mapping can be found in Fig. 11. The class diagram semantics has two variants for the mapping of super-classes. The variant `mapSuperCDirect` carries an additional constraint which excludes the use of variant `SingleInheritance` for the system model. The actual implementation of the theories has been omitted.

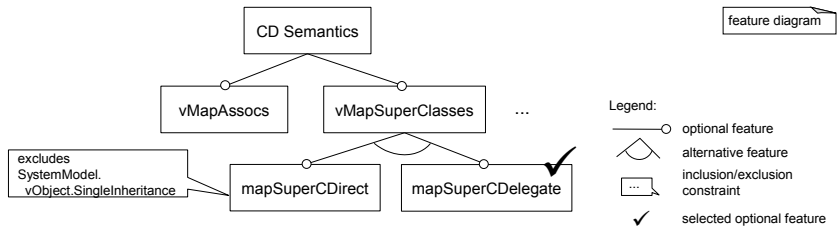

**Fig. 11.** Semantic mapping feature diagram (fragment)

### 4.3   Configuration

The configuration space of the simple class diagram language has been defined above with the help of feature diagrams. A concrete configuration for a system model is also given in Fig. 10, in which the single inheritance variant is selected as indicated by the check mark. As a configuration for the class diagram semantic mapping, we select variant `mapSuperCDelegate` (see Fig. 11); choosing the other variant would violate the exclusion constraint.

A generator written for MontiCore processes a set of configuration files (multiple configurations of, e.g., the system model may be defined). It first combines configuration files that refer to the same feature diagram. Then, it checks if the configurations conform to the feature diagrams and if the constraints have been observed. Afterwards, the configured theories for the system model and the semantic mapping are generated. In case of a system model configuration, the generated theory (see Fig. 12) combines all predicates (line 4) from the imported theories that constitute the configuration. This is done by name convention: The theory `SingleInheritance` has to provide a predicate called `valid-SingleInheritance`.

```
1  theory SystemModel imports SystemModel-base
2                              "vObject/SingleInheritance"
3  begin
4  constdefs "valid sm == valid-base sm ∧ valid-SingleInheritance sm"
5  end
```

**Fig. 12.** Resulting generated system model theory in Isabelle/HOL

Fig. 13 shows the resulting (generated) class diagram semantic mapping. It simply combines the chosen theories using the Isabelle/HOL import mechanism. The loose end in Fig. 8, namely the declaration `mSuperClasses`, is automatically bound to the definition provided in theory `MapSuperCDelegate`.

```
1  theory CDSimpSem imports CDSimpSem-base
2                              "vMapSuperClasses/MapSuperCDelegate"
3  begin end
```

**Fig. 13.** Resulting generated class diagram semantics theory in Isabelle/HOL

Finally, the theory in Fig. 14 uses the generated semantic mapping theory. The generated system model theory was already used in Fig. 8 by the base version of the semantic mapping. Presenting a meaningful verification application is outside the scope of this paper, a simple verification example has been given in [2]. The scenario in Fig. 14, however, suffices to show, on the one hand, how variants in a language definition can be systematically handled using feature diagrams. On the other, it shows that the whole approach can be supported by tools. In this scenario, property P (Fig. 14, l. 3) ranges over all class diagrams and all systems. In [2], we also presented an additional generator that translates concrete textual models to instances of the generated abstract syntax data type. This makes it also possible to reason about properties of concrete models.

The instantiation of variants is done at the theory level. We could have made all variation points type parameters, similar to [14]. A configuration would then correspond to instantiating type parameters with concrete types. We refrained from doing so because the readability of the theories would have been drastically reduced and it would be no longer possible to leave variants underspecified.

```
1  theory myVerifyApp imports CDSimpSem
2  begin
3  lemma "∀ cd sm . mCDDefinition cd sm ∧ valid sm  ⟶ P cd sm"
4    ... done
5  end
```

**Fig. 14.** A possible verification scenario in Isabelle/HOL

## 5   Related Work

To the best of our knowledge, there is no previous work on a general classification of variability mechanisms offered by modeling languages. [15] also suggests feature models to express language variabilities. The focus is on syntactic variability and variable code generators, formal semantics is not addressed.

Regarding the presented tool support for formal language definitions, most related approaches do not consider variability. For example, a complete language definition (including syntax, typing rules, and operational semantics) can be expressed in Alloy [16], which has the advantage of immediate analyzability. Semantic anchoring [17] is another approach to define semantics with tool support. Operational semantics is given by generated abstract state machines.

Other works support semantic variability to a certain extent. Template semantics [18] can be used to define the behavioral semantics of state-based modeling notations. The execution semantics is based on parametric hierarchical transition systems whose behavior can be configured with the help of predefined template parameters. In [19], template semantics is employed to define the semantics of UML state machines. The semantics explicitly models the variability found in the UML standard. [20,21] describe semantically configurable Java code generation and analyzable models using template semantics. Template semantics provides a rich theory for state-based modeling notation variants but is restricted to behavioral semantics that furthermore fits the computational model. Templatable metamodels, introduced in [22,23], is a similar approach presented for metamodeling the abstract syntax and operational semantics of a domain specific modeling language. It uses the UML 2 profile and template mechanisms to define variation points at the metamodel level and to bind the introduced generic types to concrete types at the metamodeling or modeling level. Like template semantics, the approach is targeted towards behavioral semantics but its mechanisms are more compliant with the UML standard. Quite differently, [24] proposes an approach to model semantic variation points and implementation choices as class models in their own right. These are transformed together with a source UML model into a specific target UML model that reflects the chosen variants. The focus in this work is also behavioral semantics in that variants correspond to operations implemented in an action language. We are not aware of any other framework that supports defining and configuring syntactic and semantic variability in a formal language definition.

## 6  Conclusion

The contribution of this work is twofold. First, we presented a taxonomy of variability mechanisms that may be found in a modeling language definition. Variability may be of presentation, syntactic or semantic nature. Opposed to UML, which only talks about "semantic variation points" in general, we further classify semantic variability according to semantic domain and semantic mapping variability. Semantic domain variability can be thought of as variability in some run-time system modeling the underlying platform assumptions, while semantic mapping variability would correspond to configuration options in a generator targeting a previously chosen (i.e., configured) run-time system.

Second, we extended our framework for defining the syntax and semantics of an object-oriented modeling language by integrating the variability mechanisms that we have identified. The tool suite built on MontiCore and Isabelle/HOL uses feature diagrams with inclusion/exclusion constraints to model variants and their interdependencies in the syntax, semantic domain, and semantic mapping. Given a configuration of variants for possibly multiple modeling languages, the framework generates a set of theories representing the integrated language definitions. This set of theories can be used in several verification scenarios. Note that, while the framework is tailored towards object-oriented modeling languages, the taxonomy mentioned above applies to any kind of modeling language. Likewise, the framework could be used for semantic domains other than the system model.

Future work will be concerned with elaborating variability for concrete modeling languages; larger case studies will contribute to validate the proposal and, in particular, the tool support. The long term goal, regarding one of our main targets UML, is to provide a comprehensive feature model for UML variability which ultimately could replace the currently used informal definitions and feature support statements. Another line of work is verification within our framework. Theorem proving is challenging. The effect of variability in concrete verification scenarios is not very well discussed and may require substantial further research.